\documentstyle[12pt]{article}
\def\agt{\stackrel{{\textstyle>}}{\sim}}
\def\alt{\stackrel{{\textstyle<}}{\sim}}

\def\gg{\gamma \gamma}
\def\gamgam{\gamma \gamma}
\def\hh{H^+H^-}
\def\ww{W^+W^-}

\def\nutau{\nu \tau^+ \bar\nu \tau^-}
\def\cs{c\bar s \bar c s}
\def\qq{q\bar q q\bar q}
\def\mixed{\nu \tau^+ \bar{c} s,~ c \bar{s} \bar{\nu} \tau^-}
\def\tautau{\tau^+\tau^-}
\def\branchnutau{Br$(H^+\to \nu \tau^+)$}
\def\branchcs{Br$(H^+\to c\bar s)$}
\def\branchtau{Br$(\tau\to \nu_{\tau} + \pi,\rho,a_1)$}
\def\Hmix{H^+ H^- \to \nu \tau^+ \bar{c}s,~c \bar{s} \bar{\nu} \tau^-}
\def\tHp{\theta}
\def\sqee{\protect\sqrt{s_{ee}}}

\def\epstau {\epsilon_\tau}
\def\evistau {E^{{\rm vis}}_\tau}
\def\mhp {m_{H^{\pm}}}

\def\tenfb{10~{\rm fb}^{-1}}

\def\beq{\begin{equation}}
\def\eq{\end{equation}}

\begin {document}

\begin{titlepage}

\hfill{NUHEP-TH-93-7}

\hfill{CPP-93-6}

\hfill{UTTG-8-93}

\LARGE

\vspace{0.08in}

\begin{center}
Detecting an Intermediate Mass Charged Higgs \\[.1in]
at $\gamma \gamma$ Colliders
\vspace{0.18in}
\end{center}

\normalsize

\begin{center}

David Bowser-Chao\footnote{Research supported in part by DOE Grant
DOE-FG03-93ER40757 and the Texas Advanced Research Program.}\\
{\it Center for Particle Physics, University of Texas at Austin}\\
{\it Austin, Texas 78712, USA}\\

\vspace{.15in}

Kingman Cheung\footnote{Research supported in part by DOE Grant
DOE-FG02-91ER40684.}\\
{\it Department of Physics \& Astronomy, Northwestern University}\\
{\it Evanston, Illinois 60208, USA}\\

\vspace{.15in}

Scott Thomas\footnote{Research
supported in part by the Robert A. Welch Foundation
and NSF Grant PHY 9009850.}\\
{\it Theory Group, Physics Department, University of Texas}\\
{\it Austin, Texas 78712, USA}\\

\vspace{0.25in}

\baselineskip=20pt

Abstract
\end{center}

The detection of an intermediate mass charged Higgs boson
at $\gamgam$ colliders via the modes $\gamgam \to H^+ H^-
\to \nu \tau^+ \bar{\nu} \tau^-,~ c \bar{s} \bar{c} s,~$ and $
\nu \tau^+ \bar{c}s + c \bar{s} \bar{\nu} \tau^-$ is considered.
$W^+ W^-$ boson pair production is the dominant background
for these modes. The three modes may be used in a complementary
fashion to detect a charged Higgs boson. The mixed
leptonic-hadronic mode may be used to determine the charged
Higgs mass by reconstructing the invariant hadronic mass.
The sensitivity of Br$(H^+ \rightarrow c\bar
s,\,\nu \tau^+)$ on the discovery limit of the charged Higgs boson
is also discussed.

\end{titlepage}

\baselineskip=24pt

\section{Introduction}
\label{intro}

The symmetry breaking sector of the standard model will be a
prime target of future colliders. Any enlargement of the sector beyond the
single $SU(2)_L$ Higgs doublet of the minimal standard model necessarily
involves new physical particles. With two or more doublets, as required in
supersymmetric theories, the physical spectrum includes charged Higgs
bosons. Technicolor theories can also lead to fairly light charged
technipions. In this letter the production and detection of
such charged scalars
(subsequently referred to as charged Higgs bosons)
in the intermediate mass range $m_W \alt \mhp \alt 2 m_W$
at proposed high energy $\gamgam$ colliders \cite{ggcoll}
will be considered.

Assuming the Higgs-matter coupling is proportional to mass,
the most promising means of production in hadronic colliders
is by model-dependent associated production with, or decay of,
the top quark;
e.g., $t \to b H^+$ for $\mhp < m_t - m_b$ \cite{ssc}.
At $e^+e^-$ colliders charged Higgs bosons are pair produced via
$s$-channel $\gamma$ and $Z$ exchange \cite{lep200}.
At $\gamgam$ colliders pair production proceeds through the model-independent
$H^+ H^- \gamma$ gauge coupling, as shown in Fig.~\ref{feyn}.
The primary advantage of $e^+e^-$ and
$\gamgam$ colliders with respect
to hadronic colliders is of course the
relative paucity of backgrounds for hadronic $H^{\pm}$ detection modes.

The dominant decay modes depend on the charged Higgs mass, $\mhp$.
For an intermediate mass charged Higgs boson, the decay mode
$H^+ \to Z W^+$ is not available.
With the current limit on the
lightest neutral higgs mass, $m_{h^0} > 48$~GeV \cite{h0limit}, the
mode $H^+\to h^0 W^+$ is closed for $\mhp \alt 130$~GeV. If LEP~II
pushes the bound on $m_{h^0}$ to its range of detectability
(about $m_W$), this channel is also closed for $\mhp < 2m_W$.
The mode $H^+ \to \gamma W^+$ is absent at tree level and
should be correspondingly suppressed.
For $\mhp < m_t+m_b$, where the top
quark mass is assumed to be
$m_t \approx 150$~GeV, the mode $H^+\to t \bar b$ will
also be closed. With these assumptions, the dominant decay modes
are $H^+ \to \nu\tau^+,c\bar s$.
The decay $H^+ \to c \bar{b}$ is suppressed by the small
mixing between second and third quark generations.
Decays to other quarks and leptons are
suppressed by small masses.
To this extent, \branchnutau$~+~$\branchcs~ $ \simeq 1$
over the mass range considered here.

The three decay modes of the charged Higgs pair,
$\hh\to \nutau,~\cs,$ and $\nu\tau^+ \bar{c} s +
c \bar{s} \bar{\nu} \tau^-$ may be used
in a complementary
fashion to cover the whole intermediate mass range. The last decay mode, where
detectable, may be used to determine $\mhp$ by reconstructing
the invariant mass of the $cs$ system.

\section{Production of Charged Higgs Bosons}
\label{prod}

The initial state
photons of the $\gamgam$ collider
can be produced by the laser back-scattering
method \cite{ggcoll} at a next-generation linear $e^+e^-$
or $e^-e^-$ collider.
Compton scattering of laser photons (with an energy $\omega_0$ of a
few eV) head-on with electron or positron beams of energy $E_0$
produces
back-scattered photons with energy $\omega = x E_0$.
The $e \gamma \to e \gamma$ conversion efficiency is taken to be
100\%, multiple scattering is ignored, and the incoming photons
are taken to be unpolarized.
Under these assumptions, the
back-scattered photon luminosity function
is given by \cite{ggcoll}
\beq
F_{\gamma /e}(x) = \frac{1}{D(\xi)} \left[ 1-x +\frac{1}{1-x}
-\frac{4x}{\xi(1-x)} + \frac{4x^2}{\xi^2 (1-x)^2} \right] \,,\label{lumin}
\eq
where $D(\xi)$ is a normalization factor,
\beq
D(\xi) = (1-\frac{4}{\xi} -\frac{8}{\xi^2}) \ln(1+\xi) + \frac{1}{2} +
\frac{8}{\xi} - \frac{1}{2(1+\xi)^2}\,,
\eq
and $\xi=4E_0\omega_0/m_e^2$.
The incoming photon energy, $\omega_0$, is chosen such that
$\xi=2(1+\sqrt{2})$ to maximize the back-scattered photon energy
while avoiding $e^+e^-$ pair creation.
The photon luminosity function
vanishes for $x > x_{\rm max}=\xi/(1+\xi) \simeq 0.83$.
For definiteness the design parameters of
the Next Linear Collider (NLC),
with center of mass (CM) energy $\sqrt{s_{ee}} = 500$ GeV and luminosity
${\cal L}_{ee}$ = 10 fb$^{-1}$ yr$^{-1}$ will be employed.
The mass range of a
charged Higgs search in the $\gamgam$ mode at the NLC
is limited to $\mhp < x_{\rm
max} \sqrt{s_{ee}}/2 \simeq 207$~GeV, well above the intermediate mass
range considered here.

The cross section $\sigma$ for any process is the convolution of
the hard-scattering cross section $\hat{\sigma}$ with the
photon luminosity functions,
\beq
\label{convol}
\sigma(s_{ee}) = \int_{\tau_{\rm min}}^{x_{\rm max}^2}  d\tau \label{totsigma}
\int_{\tau/x_{\rm max}}^{x_{\rm max}} {dx_1 \over x_1}
F_{\gamma/e}(x_1) F_{\gamma/e}(\tau/x_1) \hat \sigma
(\hat{s}_{\gamgam} = \tau s_{ee})\,,
\eq
where
$$
\tau_{\rm min}   =   \frac{(M_{\rm final})^2}{s_{ee}}\,,
$$
and $M_{\rm final}$ is the sum of final state particle masses.

The differential
cross section for the process $\gamgam\to\hh$
(shown in Fig.~\ref{feyn}) is given by
\beq
\frac{d\hat{\sigma}}{d~|\cos \tHp|} =
\frac{ \pi \alpha^2 \beta}{\hat{s}_{\gamgam}}
\left(  1 - \frac{2 \beta^2 (1-\beta^2) \sin^2 \tHp}
                 {(1 - \beta^2 \cos^2 \tHp)^2} \right)
\eq
where $\hat{s}_{\gamgam}$ is the photon pair CM mass energy, $\tHp$ is
the $\gamma - H^+$ CM polar angle, and
$\beta \equiv \sqrt{1 - 4 \mhp^2 / \hat{s}_{\gamgam}}$.
For
$\sqrt{\hat s}_{\gamgam} >> \mhp$, $\hat{\sigma}$ falls as $s^{-1}_{\gamgam}$
(as expected), and becomes isotropic in the CM frame.
This is helpful,
as the principal backgrounds from pair production of fermions and
spin-one gauge bosons,
while larger in magnitude,
are sharply peaked in the forward direction
for large $\hat{s}_{\gamgam}$.
The $\gamgam \to H^+H^-$ total cross section for $\sqrt{s_{ee}} = 0.5$ TeV
and 1 TeV, including the convolution (\ref{convol}),
are shown in Fig.~\ref{hhfig}.
For small Higgs mass the cross section is actually larger
for 0.5 TeV than for 1 TeV.
This is because the $\hat{s}_{\gamgam}$ dependence convoluted with the
soft portion of the photon luminosity function
increases the $H^+H^-$ production rate for
$\mhp << x_{\rm max} \sqrt{s}/2$, compared with the monochromatic
case.

Before considering detection of the charged Higgs via the three modes,
it is necessary to
comment on the assumptions made for detection of decay products,
and on the acceptance cuts employed.
For $\tau^{\pm}$ identification, the leptonic decay modes suffer
irreducible backgrounds from $W^{\pm}$ decay discussed below.
Only the hadronic modes $\tau^- \to \pi^-  \nu_\tau,~\rho^-
\nu_\tau, ~a_1^- \nu_\tau$,  with a total branching ratio 0.45,
will be considered.
Requiring visible decay product energy $\evistau > 10$ GeV ensures that
the decay products travel essentially along the original $\tau^{\pm}$
direction.
The $\tau^{\pm}$ decay product distributions depend on the $\tau^{\pm}$
charge, helicity, and decay mode.
Summing over the hadronic modes given above,
the decay product laboratory energy distribution may
be approximated by \cite{tsai,taudecay}
\beq
\frac{1}{\Gamma} \frac{d \Gamma}{d z} =
     0.45  - 2SC (0.43 z - 0.215)
\label{ztau}
\eq
where $C$ and $S$ are the $\tau^{\pm}$ charge and helicity,
$z=\evistau / E_{\tau}$,
$E_{\tau}$ and $\evistau$ are the $\tau^{\pm}$ and decay product laboratory
energies.
The decay products of $\tau^{\pm}$ coming from $H^{\pm}$ decay
have a somewhat harder distribution than from $W^{\pm}$ decay
\cite{taudecay} (the main background discussed below).
The $\evistau > 10$ GeV cut therefore enhances the signal to
background ratio.
This ratio could be further enhanced by using correlations among the
multi-pion final states \cite{taudecay}.

In the absence of definite detector designs,
representative parameters will be employed \cite{workshop}.
The hadronic calorimeter is taken to have gaussian resolution
with standard deviation
\beq
\frac{\delta E}{E} (\%) = \sqrt{\frac{a^2}{E} +b^2}
\label{resolution}
\eq
with $a=60$ GeV$^{1/2}$ and $b=2$.
Calorimeter and
tracking coverage are assumed to extend
to $|\cos\theta| < 0.95$.
Photon identification is assumed to extend to $|\cos\theta|< 0.985$.
Aside from the
$\tau^+ \tau^- \gamma$ background discussed in the next section,
all visible decay products in the laboratory frame are required
to have $|\cos \theta| <$ 0.95,
be separated by at least
$15^{\circ}$,
and have an energy
$E^{\rm vis} > 10$ GeV.
In addition to allowing for realistic detector acceptances, these
cuts help suppress the backgrounds discussed below.
Quark jet identification is taken to be 100\%, but
explicit dependence on
$\tau^{\pm}$ identification efficiency, $\epstau$, is retained below.

\section{$\hh \to \nutau$}
\label{mode1}

The first decay mode,
\beq
\gamgam \to \hh \to \nutau
\label{first}
\eq
has  background from the following processes:
\begin{eqnarray}
\gamgam &\to ~\tautau & \label{1:tautau}\\
\gamgam &\to ~\ww &\to ~\nutau \label{1:ww} \\
\gamgam &\to ~\tautau Z &\to ~\tautau \nu \bar\nu \label{1:tautauZ}\\
\gamgam &\to ~\tau^+ \tau^- \gamma & \label{1:tautaugam}
\end{eqnarray}
 The total cross section
from process~(\ref{1:tautau}) is daunting at 49~pb
(including the $\tau$ branching ratios). Because of $t$-channel
enhancement though,
the $\tau$'s are sharply peaked in the forward direction.
The $|\cos\theta_{\tau}| < 0.95$ cut
reduces the cross section to 9.2~pb. Since this
is a two-body final state, however,
the
azimuthal angle $|\Delta\phi|$ between the $\tau$'s will be $180^\circ$.
The requirement $\evistau > 10$~GeV
sufficiently collimates
the decay products along the original $\tau^{\pm}$ directions so that
a cut of $|\Delta\phi| < 170^\circ$ should eliminate this background.
To the extent that the $\hh$ pair is not extremely relativistic,
the $\tau$'s from (\ref{first}) have a relatively uniform distribution
in $\Delta\phi$. An even stronger acoplanarity cut, if required, would
therefore not significantly reduce the signal.

Process~(\ref{1:ww}) involves
the small branching ratio [Br$(W^+\to\nu\tau^+)]^2 \simeq (1/9)^2$.
The $W^+W^-$ pair is
largely transversely polarized  \cite{WWback} and peaked strongly in the
forward-backward direction.
The $\tau$'s therefore tend to be softer or boosted along the beam.
As
discussed above, the decay products of a $\tau^{\pm}$
coming from $W^{\pm}$ decay
have a softer distribution
than those from $H^{\pm}$ decay, thus aiding the efficiency of the
$\evistau$ cut. The acoplanarity cut does not greatly affect this
background.
After all cuts this background is 40 fb.

Although processes~(\ref{1:tautauZ}--\ref{1:tautaugam})
are higher order, the extra particle(s) in the final state
potentially reduce the effectiveness of the acoplanarity cut.
Both have a $t$-channel
enhancement though, that is suppressed
by the $|\cos\theta_{\tau}| < 0.95$ cut.
This effectively eliminates the
background (\ref{1:tautauZ}),
which after all cuts is only 0.26 fb.
The background (\ref{1:tautaugam})
is greatly reduced by vetoing on the presence of a ``visible''
photon, defined as being within the detector
($|\cos\theta_{\gamma}| < 0.985$),
having energy greater than 10~GeV, and being farther
than $5^\circ$ from either $\tau$.
What is left of this process requires that the photon be
along one of the $\tau$'s, go down the
beam, or be fairly soft in the central region.
In any of these cases, the azimuthal angle between the $\tau$'s
should not be very different from $180^\circ$, so that the
acoplanarity cut effectively reduces this background.
After all cuts this background is 5.6 fb.

The cross section for the signal (\ref{first}) as a function of
$\mhp$,
after all cuts, is shown in Fig.~\ref{mode1fig}.
It ranges from
79~fb (for $\mhp = 50$~GeV) to 10.6~fb (for $\mhp = 150$~GeV), assuming
\branchnutau = 1. The backgrounds from processes~(\ref{1:ww}) and
(\ref{1:tautaugam}) are also shown.
In this figure, the $\tau$ detection efficiency,
$\epsilon_{\tau}$, is taken to be
100\%.
The actual cross sections of course scale like $\epsilon_{\tau}^2$.

\section{$\hh\to \cs$}
\label{mode2}

The second decay mode
\beq
\gamgam \to \hh \to \cs\,,
\eq
has the principal background
\beq
\gamgam \to \ww \to \qq\,. \label{2:ww}
\eq
In contrast to the analogous process of the previous section,
this background is not greatly reduced by the branching
ratio [Br$(W^{\pm} \to q \bar{q})]^2 \simeq (2/3)^2$.
With all the cuts discussed in section~\ref{prod}, this
background amounts to 12 pb.
This may be reduced by vetoing  events for which the dijet
invariant mass,
$m_{q_i q_j} \simeq \sqrt{2 p_{q_i} \cdot p_{q_j} }$,
of any pair of jets, roughly reconstructs
the $W^{\pm}$ mass.
Including the hadronic calorimeter resolution (\ref{resolution})
and the cut
$|m_{qq} - m_W| >$ 8 GeV on all jet pairs results in a reduction
of a factor of approximately 80, to 0.1517 pb.
Thus about 10\% of all on-shell $W^{\pm}$ decays remain after
this cut.
The entire background is now due to the hadronic calorimeter smearing
of the two on-shell $W^{\pm}$ decays.
It is therefore worth considering whether the next order tree
level electroweak processes (see Fig.~\ref{qqWfig}) contained in
\beq
\gamgam \to q\bar q W \to \qq  \label{2:qqw}
\eq
contribute substantially since only one of the $qq$ pair invariant
masses necessarily reconstructs $m_W$ (before including hadronic calorimeter
resolution). The cross section for (\ref{2:qqw}) with all cuts (including
that on $m_{qq}$) is 0.1533 pb, almost identical to the
cross section from on-shell $W^+W^-$ decay.
This result is not surprising  since the full gauge invariant set of
Feynman diagrams for (\ref{2:qqw}) is dominated by a subset
of diagrams, identical to the ones contributing to (\ref{2:ww}),
in which one of the $W^{\pm}$ is {\it nearly} on-shell.
With the energy resolution considered here the
corrections from higher order the processes contained
in (\ref{2:qqw}) are therefore
unimportant.
Higher order electroweak processes contained in
$\gamgam \to q \bar{q} q \bar{q}$ would give an even smaller correction.
The QCD processes $\gamgam \to q \bar{q} q \bar{q}, ~q \bar{q} g g$
would also contribute at roughly the same order as the
correction from (\ref{2:qqw}), and are therefore unimportant.
If the background from ({\ref{2:ww}) were reduced
with significantly better hadronic resolution, the higher order
processes might
become important.
Also for $\mhp$ significantly different than $m_W$ the
cut on $|m_{qq} - m_W|$ could be increased, thereby reducing
the background without greatly affecting the signal.
The choice of 8 GeV represents a compromise for the range
of $\mhp$ considered here.

Fig.~\ref{mode2fig} shows the cross sections for the signal
with acceptance cuts
as a function of $\mhp$,
with and without the requirement $|m_{qq}-m_W| > 8$~GeV.
The cross sections for the background (\ref{2:ww}), with and without the
$m_{qq}$ cut are  also shown.

\section{$\hh\to \mixed$}
\label{mode3}

The third decay mode
\beq
\gamgam \to \hh \to \mixed\,,
\label{3hh}
\eq
has the principal background
\beq
\gamgam \to \ww \to \nu \tau^+ q \bar{q},~
                     q \bar{q} \bar{\nu} \tau^-
  \,. \label{3:ww}
\eq
This mode offers the possibility of reconstructing $\mhp$ through
$m_{qq}$.
For $m_{qq}$ significantly different than $m_W$, the background
(\ref{3:ww}) contributes only through the hadronic calorimeter smearing
of on-shell $W^{\pm}$ decays.
As in the previous section, it is therefore worth considering
the next order tree level electroweak processes contained in
\beq
\gamgam \to q\bar q W^{\pm}
   \to q \bar{q} \nu \tau^+,~
   q \bar{q} \bar{\nu} \tau^- . \label{3:qqw}
\eq
The differential cross section, $d \sigma / d m _{qq}$, with
acceptance cuts,
is shown in Fig.~\ref{mode3dSigma} as a function of $m_{qq}$
for the signal (\ref{3hh}) and background (\ref{3:qqw}).
This figure assumes $\epsilon_{\tau}=1$,
\branchnutau = \branchcs = $1/2$, and the
hadronic calorimeter resolution (\ref{resolution}).
Again, the total cross section for (\ref{3:qqw}) is dominated
by diagrams identical to the ones contributing to (\ref{3:ww})
in which one of the $W^{\pm}$ is nearly on-shell.
But with the hadronic calorimeter resolution assumed, the higher-order
corrections contained in (\ref{3:qqw}) do become important for
$m_{qq} \alt 60$ GeV and $m_{qq} \agt 100$ GeV.

As evident in Fig.~\ref{mode3dSigma}, for $\mhp$ sufficiently
different from $m_W$, the background can be reduced by rejecting
events outside a window centered on $\mhp$.
Implementing this requires an $\mhp$ dependent analysis.
In order to ensure that most of the signal is retained,
a window of $\pm ~8$ GeV will be used.
The cross section for the signal (\ref{3hh}) and background
(\ref{3:qqw}) as a function of the presumed charged Higgs mass is
shown in Fig.~\ref{mode3total}.
This figure includes the same cuts and assumptions
as Fig.~\ref{mode3dSigma}, and
the requirement $|m_{qq} - \mhp| < 8$ GeV.

\section{Sensitivity to \branchnutau ~and \branchcs}
\label{results}

In this section,
the sensitivity of each mode to
\branchnutau, \branchcs, $\mhp$, and the $\tau^{\pm}$ identification
efficiency, $\epstau$,
is analyzed.
For a 5 standard deviation measurement,
the requirement
$N_s / \sqrt{N_b} > 5$ is imposed, where $N_s$ and $N_b$ are the
number of signal and background events, respectively,
including branching ratios and efficiencies.

Detection of the mode $\hh\to\nutau$ over the $\ww$ and $\tautau\gamma$
backgrounds requires
\beq
\hbox{\branchnutau} > \sqrt{
 \frac{5 \sqrt{\epstau^2 L \sigma_b}}
 {\epstau^2 L \sigma_s}}
  \label{mode1limit}
\eq
where $L= \int {\cal L}~dt$ is the integrated luminosity.
The signal and background cross sections, $\sigma_s$ and
$\sigma_b$, are from Fig.~\ref{mode1fig}.
The upper curve in Fig.~\ref{branchfig}
gives
the minimum value of \branchnutau~ accessible in this mode,
for $L= \tenfb$, and
optimistically assuming $\epstau=1$ after all cuts.
A smaller value of
$\epstau$ could be offset by an increase in $L$.
In addition, increasing the cut $\evistau>10$~GeV
to $\evistau > 20$~GeV leaves the ratio $\sqrt{\sigma_b}/\sigma_s$
approximately constant over the range of $\mhp$ considered.
This
would increase $\epstau$ by improving tracking efficiency for
hadronic $\tau^\pm$ decays.

The analogous analysis for the mode $\hh\to\cs$ requires
\beq
\hbox{\branchcs} >
   \sqrt{\frac{5\sqrt{L \sigma_b}}{L\sigma_s}}
\label{mode2limit}
\eq
where $\sigma_s$ and $\sigma_b$
are from Fig.~\ref{mode2fig}.
For $| \mhp - m_W | \agt 8$ GeV
the $m_{qq}$ cut  discussed in Section ~\ref{mode2} is imposed,
while for $| \mhp - m_W | \alt 8$ GeV better significance is obtained
without the $m_{qq}$ cut.
The minimum value of \branchcs~ accessible in this mode for
$L = \tenfb$ is shown in Fig.~\ref{branchfig}.

Detection of the mixed mode $\Hmix$ requires
\beq
\min\left(\hbox{\branchnutau},\hbox{\branchcs}\right)
     >
\frac{1}{2}
-\frac{1}{2}
 \sqrt{1-
 \frac{5 \sqrt{\epstau L \sigma_b}}
 {\epstau L \sigma_s}}
\label{mode3limit}
\eq
where $\sigma_s$ and $\sigma_b$ are from Fig.~\ref{mode3total} and
\branchnutau $+$ \branchcs $=1$ is assumed.
The lower curve in Fig.~\ref{branchfig}
gives the minimum branching ratio accessible
in this mode for $L = \tenfb$ and $\epstau = 1$.
This minimum branching ratio grows as
$1/\sqrt{\epstau}$ for $\mhp\agt100$~GeV.

As expected, the three modes considered
here are complementary in the search for an intermediate mass
charged Higgs.
The mixed mode
$\Hmix$
offers the most
complete coverage for discovery, assuming a reasonable $\tau$
identification efficiency and that neither branching ratio is
very small. The other two modes may be used for confirmation
over a somewhat more limited region of parameter space. If one
of the branching ratios is too small for the mixed mode,
a charged Higgs could be detected in the remaining mode.
The entire intermediate mass range for the charged Higgs boson
can therefore be covered by the NLC in the $\gg$ mode.

\section{Acknowledgements}
\label{ack}

We would like to thank D. Dicus, C.~Kao, and K.~Lang for useful discussions.
This research was supported in part by DOE Grants
DOE-FG02-91ER40684 (K. C.), and DOE-FG03-93ER40757 (D. B.-C.),
the Texas Advanced Research Program (D. B.-C.),
the Robert A. Welch Foundation (S. T.)
and NSF Grant PHY 9009850 (S. T.).
Computing resources were provided in part by the University
of Texas Center for High Performance Computing.

\clearpage
\begin{figure}[]
\caption{\label{feyn}
Feynman diagram for $\gamma \gamma \to H^+ H^-$.
The crossed and seagull diagrams are not shown.
}
\end{figure}
\begin{figure}[]
\caption{\label{hhfig}
Cross section (pb) for $\gamgam\to\hh$ versus $\mhp$
for $\sqee$ = 0.5 TeV and 1 TeV.
}
\end{figure}
\begin{figure}[]
\caption{\label{mode1fig}
Cross section (pb) for $\gamgam\to\hh\to\nutau$ versus $\mhp$
for $\sqee$ =  0.5 TeV,
including the acceptance cuts. The branching ratio
\branchnutau~= 1, $\epsilon_{\tau}$ = 1, and
\branchtau$=0.45$ are
included. The horizontal lines are the $\ww\to\nutau$
and $\tau^+ \tau^- \gamma$
backgrounds.
}
\end{figure}
\begin{figure}[]
\caption{\label{qqWfig}
Feynman diagram for $\gamgam\to q\bar q W$ in which only {\it one}
on-shell $W$ is produced.
Other diagrams related by gauge invariance, and including other
quark flavors are not shown.
}
\end{figure}
\begin{figure}[]
\caption{\label{mode2fig}
Cross sections (pb) for $\gamgam\to\hh\to\cs$
versus $\mhp$
for $\sqee$ = 0.5 TeV,
including different sets of cuts. The lower
(upper) solid curve does (not) include
the requirement $|m_{qq} - m_{W}| >$ 8 GeV;
the branching ratio \branchcs ~= 1. The
lower (upper) dashed curve represents the $\ww\to q\bar qq\bar q$
background with (without) the
$m_{qq}$ cut.
Both include the
acceptance cuts.}
\end{figure}
\begin{figure}[]
\caption{\label{mode3dSigma}
Differential cross section $d\sigma/dm(q\bar q)$ (pb/GeV) for
$\gamgam \to \hh\to\mixed$
(in solid curves) versus $m_{qq}$
with $\mhp =$ 60, 80, 100, 120, and
140~GeV, and the background
$\ww \to\ \nu \tau^+ q \bar{q},~ q \bar{q} \bar{\nu} \tau^-$
{}~(dashed curve)
for $\sqee$ = 0.5 TeV;
$\epstau = 1$, \branchnutau~=~\branchcs~=~0.5, \branchtau = 0.45,
and acceptance cuts included.
}
\end{figure}
\begin{figure}[]
\caption{\label{mode3total}
The total cross section (pb) for $\gamgam \to \hh\to\mixed$ (solid
curve) and the background
$\ww \to \nu \tau^+ q \bar{q},~ q \bar{q} \bar{\nu} \tau^-$
{}~(dashed curve) versus
the presumed Higgs mass
for $\sqee$ = 0.5~TeV;
$\epstau = 1$,~\branchnutau~=~\branchcs~=~0.5, \branchtau~=~0.45,
$|m_{qq} - m_{H^{\pm}}| < $ 8 GeV,
and acceptance cuts included.
}
\end{figure}
\begin{figure}[]
\caption{\label{branchfig}
Minimum values of \branchnutau~ and \branchcs~
 for the given decay mode to have
significance of at least 5, assuming an
integrated luminosity of 10~fb$^{-1}$ and $\epsilon_{\tau} = 1$.
}
\end{figure}

\end{document}